# The light curve and the time delay of QSO 0957+561

**J. Pelt**[1], **R. Kayser**[2], **S. Refsdal**[2], and **T. Schramm**[2]

[1] Tartu Astrophysical Observatory, EE-2444 Tõravere, Estonia
[2] Hamburger Sternwarte, Gojenbergsweg 112, D-21029 Hamburg-Bergedorf, Germany

**Abstract.** We present a new analysis of the presently available photometric data for the gravitationally lensed quasar 0957+561 A,B with the aim of determining the time delay between its two images. The basic method used is the dispersion estimation technique. Even by using the simplest non-parametric form of our method we can convincingly rule out a time delay near 536 days and show that the time delay is in the vicinity of 420 days. We then introduce refinements to our method in order to get a stable and reliable result for the time delay independent of high frequency noise in the data and sampling errors. Our best result for the time delay, checked by various statistical tests and using bootstrap error estimates, is 423±6 days. We furthermore confirm our earlier result that the radio data are compatible with this value. Using the best available model for the mass distribution in the lensing galaxy and cluster, our result for the time delay constrains the Hubble parameter to be smaller than 70 km/(s Mpc).

## 1. Introduction

The time base of the optical light curves of the double quasar QSO 0957+561A,B has been significantly enhanced due to observations performed mainly by R. Schild and his collaborators (Schild & Thomson 1994). An analysis of the much smaller data set presented by Vanderriest et al. (1989) as well as the radio data of Lehár (1992) led Press et al. (1992a,b) to an estimation of the time delay between the images of 536 days. Reanalysing the data with a different but nevertheless equally sound statistical method convinced us (Pelt et al. 1993, hereafter Paper I) that the result of 536 days is in a statistical sense rather unstable and that a time delay around 410 days is at least as probable as the delay obtained by Press and his collaborators. We present here an analysis of the presently available data with refined versions of our new methods as described in Paper I. Our result not only gives the first statistical reliable estimation of the time delay but also gives evidence for gravitational microlensing in at least one of the two light curves.

The time delay between the images of a gravitationally lensed object is of great astrophysical interest since it may be used to determine the Hubble parameter as well as the mass of the lens (Refsdal 1964a,b, Borgeest & Refsdal 1984, Borgeest 1986, Kayser 1990, Falco et al. 1991b). The double quasar 0957+561 A,B (Walsh et al. 1979, Young 1980, Falco 1992) is up to now the only gravitational lens system for which serious attempts have been made to determine the time delay $\tau$ between its images (cf. Paper I and references therein). However, the controversy around the different results obtained by different groups of authors has not yet been decided.

After a description of the data set used throughout our paper (Sect. 2) we recall in Sect. 3.1 the methods used in Paper I and describe some refinements of the techniques which we apply later. In the following subsections we analyse the light curves with increasing accuracy, compare results to previous ones and conclude our best result for the time delay, including an estimation of the amount of microlensing.

In addition (Sect.3.5) we reanalyse the radio data (Lehár et al. 1992) with the refined methods and confirm the result of Paper I that the radio data are compatible with the result obtained from the optical data.

In Section 4 we discuss the methodology, our results and their physical interpretation and meaning, including constraints on the Hubble parameter.

## 2. Data sets

The extended data set, kindly made available to us by R.E. Schild and D.J. Thomson (1994, below ST94), contains observations from November 1979 to July 1994 (JD. 244193.9–249540.7). While it contains observations from different sources, the overwhelming majority of the data points have been measured by R. Schild and his collaborators. It is therefore the most homogeneous data set available for QSO 0957+561. A full description of the data set with explanation of reduction, correction and compilation procedures can be found in Schild and Thomson (1994).

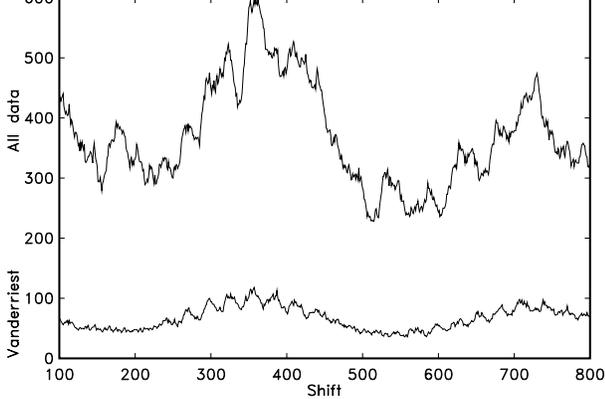

**Fig. 1.** The window function $N_{A,B}(\tau)$ for the two data sets. Upper curve: data set of Schild and Thomson (1994, ST94); lower curve: data used by Vanderriest et al. (1989, V89)

There are in total 831 time points with photometric estimates for both components of the double quasar. For the A and B light curves the full amplitude of variability is 16.36–16.96 and 16.27–16.96, respectively. The mean error for both light curves as estimated by the observers is 0.0158 magnitudes, giving a mean dispersion of $D^2 = 0.000251$. However, one should note a striking difference between the first and second part of the ST94 data: The dispersion is significantly decreasing in time.

There is a simple statistic which can be used to characterise the sequences of observations from the point of view of usefulness in the search for time shifts. The window function $N_{A,B}(\tau)$ measures the number of pairs of nearby observations in the combined data set (where the B light curve is shifted in time by $\tau$) which contain one observation from curve A and one observation from curve B (cf. Paper I). In Fig. 1 the window function $N_{A,B}(\tau)$ is shown for two different data sets–the ST94 data and the data compiled by Vanderriest et al. (1989, below V89). The absolute minimum (43 pairs) for the V89 data occurs for shifts in the range 534–537 days. This was one of the reasons which guided us in Paper I to seek another probable value for the time delay besides the 536 days proposed by Press et al. (1992a). For the ST94 data the absolute minimum of $N_{A,B}(\tau)$ occurs for shifts of 514 and 515 days with 327 pairs. Consequently the statistical reliability (even for the shifts with most unfortunate time point spacings) is now at least seven times higher.

### 3.1. Dispersion spectra

To estimate the time delay between the images A and B we use basically the same methods as described in Paper I. Nevertheless we repeat here shortly the general setup and describe some new variations of our techniques.

Our data model consists of two time series:

$$A_i = q(t_i) + \epsilon_A(t_i), i = 1, \ldots, N_A \qquad (1)$$

and

$$B_j = q(t_j - \tau) + l(t_j) + \epsilon_B(t_j), j = 1, \ldots, N_B \qquad (2)$$

where $q(t)$ denotes the inner variability of the quasar, $l(t)$ is a variability component which contains the unknown amplification ratio and an additional low frequency noise component due to hypothetical microlensing, $\tau$ is the time delay, $\epsilon_A(t)$ and $\epsilon_B(t)$ are observational errors for which crude estimates for their dispersions $\delta_A^2(t_i)$ and $\delta_B^2(t_j)$ (or corresponding statistical weights $W_A(t_i), W_B(t_j)$) are available.

For every fixed combination $\tau, l(t)$ we generate a combined light curve $C$ by taking all values of $A$ as they are and "correcting" the $B$ data by $l(t)$ and shifting them by $\tau$:

$$C_k(t_k) = \begin{cases} A_i, & \text{if } t_k = t_i, \\ B_j - l(t_j), & \text{if } t_k = t_j + \tau, \end{cases} \qquad (3)$$

where $k = 1, \ldots, N$ and $N = N_A + N_B$. The dispersion spectra

$$D^2(\tau) = \min_{l(t)} D^2(\tau, l(t)). \qquad (4)$$

can now be computed and searched for significant minima with respect to $\tau$. In Paper I we used (with different notation) a fixed $l(t) = l_0$ as unknown magnification ratio. In this paper we do the same, but consider additionally perturbations modelled by low degree polynomials. It is important to note that the parameterization of $l(t)$ is always done in the framework of the original time points (not the shifted ones).

The properties of the dispersion spectra depend strongly on the choice of the exact form for the estimates. The simplest dispersion estimate (see also Paper I)

$$D_1^2 = \min_{l(t)} \frac{\sum_{k=1}^{N-1} W_k (C_{k+1} - C_k)^2}{2 \sum_{k=1}^{N-1} W_k}, \qquad (5)$$

where the $W_k$ are statistical weights of the combined light curve data

$$W_k = W_{i,j} = \frac{W_i W_j}{W_i + W_j}, \qquad (6)$$

does not contain any free parameters except those which describe the hypothetical microlensing (the degree of the

modification of the first statistic

$$D_2^2 = \min_{l(t)} \frac{\sum_{k=1}^{N-1} W_k G_k (C_{k+1} - C_k)^2}{2 \sum_{k=1}^{N-1} W_k G_k} \quad (7)$$

where $G_k = 1$ only when $C_{k+1}$ and $C_k$ are from different images and $G_k = 0$ otherwise, measures the dispersion specifically in the overlap areas of the combined light curve. Obviously, this is the dispersion we are interested in, whereas $D_1^2$ may become strongly dominated by the dispersion of only one of the lightcurves ($A$ or $B$) in dependence of the window function (cf. Paper I).

The pairs of consecutive observations are included into the dispersion estimates $D_1$ and $D_2$ without consideration of the distance between the two time points $t_i, t_j$. In this way two observations, one of which occurs before a long gap in time and another one which occurs after this gap can make a strong contribution to the overall estimate. However, this kind of long time correlation is certainly ruled out for the red-noise like light curves of quasars. To avoid this we therefore add an additional constraint into our dispersion estimate:

$$D_3^2 = \min_{l(t)} \frac{\sum_{k=1}^{N-1} S_k W_k G_k (C_{k+1} - C_k)^2}{2 \sum_{k=1}^{N-1} S_k W_k G_k} \quad (8)$$

where

$$S_k = \begin{cases} 1, & \text{if } |t_{k+1} - t_k| \le \delta, \\ 0, & \text{if } |t_{k+1} - t_k| > \delta \end{cases} \quad (9)$$

and $\delta$ is the maximum distance between two observations which can be considered as nearby.

The number of pairs of observations which are included in the estimates $D_1 \cdots D_3$ is normally not sufficient to avoid strong noise in the corresponding dispersion spectra. Very often there are some influential time points in the observational sequence, removal of which can change the *local* minima of the spectra quite significantly. To get statistically more stable spectra we used a fourth form of the dispersion estimates which now includes not only the neighbouring pairs:

$$D_{4,k}^2 = \min_{l(t)} \frac{\sum_{l=1}^{N-1} \sum_{m=l+1}^{N} S_{l,m}^{(k)} W_{l,m} G_{l,m} (C_l - C_m)^2}{\sum_{l=1}^{N-1} \sum_{m=l+1}^{N} S_{l,m}^{(k)} W_{l,m} G_{l,m}} \quad (10)$$

where $W_{l,m}$ are statistical weights, $G_{l,m}$ selects pairs according to the origin of the values $C_l$ and $C_m$ (as for the estimates $D_2^2$ and $D_3^2$), and $S_{l,m}^{(k)}$ weights the squares depending on the distance between corresponding time points. In the various calculations we used three forms for the weights $S_{l,m}^{(k)}$. The first one

$$S_{l,m}^{(1)} = \begin{cases} 1, & \text{if } |t_l - t_m| \le \delta, \\ 0, & \text{if } |t_l - t_m| > \delta \end{cases} \quad (11)$$

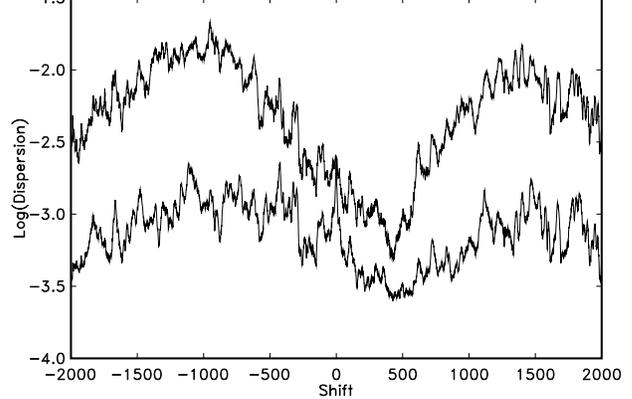

**Fig. 2.** Spectra $D_1^2$ (lower curve) and $D_2^2$ (upper curve) for the ST94 data set

includes only pairs with less distance between two observations (in the *combined* light curve) than $\delta$ (let us call this parameter *decorrelation length*). In the second scheme the pairs with longer distances between observations get linearly decreasing weights

$$S_{l,m}^{(2)} = \begin{cases} 1 - \frac{|t_l - t_m|}{\delta}, & \text{if } |t_l - t_m| \le \delta, \\ 0, & \text{if } |t_l - t_m| > \delta \end{cases}. \quad (12)$$

The third weighting scheme

$$S_{l,m}^{(3)} = \frac{1}{1 + \left(\frac{t_l - t_m}{\beta}\right)^2} \quad (13)$$

where $\beta$ plays the role of the decorrelation length, includes *all pairs* of observations into the $D_{4,3}^2$ estimate. However, in our actual calculations we were forced to use a certain cut-off distance to make computing times acceptable.

The number of different dispersion estimates gives us some flexibility when analysing the actual data but also involves a degree of arbitrariness. We therefore tried to be as conservative as possible when using different algorithms. We present our results step by step and invoke modifications only when the motivation to do so is quite obvious.

Error bars for the delay estimates were computed as described in Paper I. We applied an adaptive median filter (see Appendix) to smooth the combined light curve and reshuffled the computed residuals 1000 times to generate bootstrap samples. The shifts obtained in this way were then used to calculate the error bars.

*3.2. Standard estimates*

In this section we the describe the overall behaviour of the simplest dispersion spectra, computed for the ST94 data set. We must stress here that in the following computations the data were used without any preprocessing. No detrending or censoring was applied. Also, the hypothetical microlensing was ignored ($l(t) = l_0$).

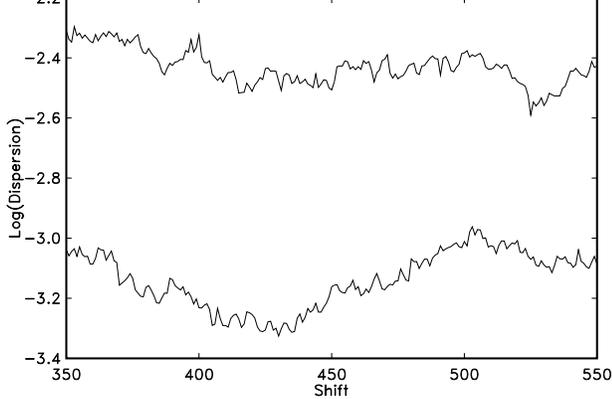
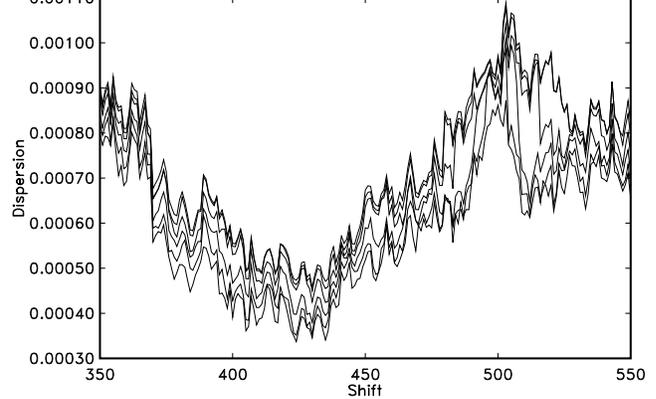

**Fig. 3.** Spectra $D_2^2$ for the ST94 data set (lower curve) and for the V89 data set used in Paper I (upper curve)

**Fig. 4.** Spectra $D_3^2$ for $\delta = 2, 4, 8, 16, 32, 64$

To get a general idea about the behaviour of the dispersion spectra for a wide range of $\tau$ values we computed $D_1^2$ and $D_2^2$ dispersion spectra (which were also used in Paper I and which do not contain any free parameters) for $\tau = -2000, -1999, \ldots, 1999, 2000$ days. In Fig. 2 the spectra $D_1^2$ and $D_2^2$ are shown in logarithmic scale to demonstrate their essential similarity.

We can see that there are essentially global minima around $400 \cdots 450$ days in both spectra. The feature around 536 days is also well seen, however it is much less pronounced than for the V89 data (see Paper I). Below we restrict the plot range for the $\tau$ values to $350, 351, \ldots, 549, 550$ days to demonstrate the behaviour of the different spectra in more detail for this important subinterval. However, the actual computations were carried out for a wider interval (typically $100, 101, \ldots, 799, 800$ days).

In Fig. 3 the fragment of the $D_2^2$ spectra is plotted with the analogous fragment computed for the V89 data set (1989). The fluctuating nature of both spectra is quite obvious. Our goal in Paper I was to demonstrate that besides the feature around 536 days, which is quite unstable against detrending and censoring applied to the V89 data, there are also other possible and plausible values for the time delay.

In the spectrum for the ST94 data the absolute minimum is at a time delay of 430 days *even without any detrending and censoring applied to the data*. Around this value, however, there are a number of other local minima (at 405, 411, 416, 424, 435 and 439 days). Because the aim of the current paper is to determine the time delay for the double quasar 0957+561 *as exactly as possible* we can not simply use the absolute minimum as the best estimate because it depends on minute details of data spacing and observational errors. To get statistically more stable estimates we are forced to introduce some additional considerations and free parameters into our estimation scheme.

The simplest and most obvious improvement is the introduction of an upper limit for the distance between data points to be included into the cross sums (estimate $D_3^2$). This excludes possible fluctuations due to the long gaps in the combined light curve.

In Fig. 4 we have plotted a sample of $D_3^2$ spectra with different values of $\delta$. There are some minor details which change from one $\delta$ value to another, but the general fluctuating nature of the spectra unfortunately remains.

### 3.3. Refinement

As we saw the behaviour of $D_1^2$, $D_2^2$ and also $D_3^2$ type dispersion spectra is somewhat erratic due to the large sampling errors (the maximum possible size for a subsample for $D_1^2, D_2^2$ and $D_3^2$ is $N - 1$ but is generally less due to the spacing of the $A$,$B$ or $B$,$A$ type pairs). It is enough to remove only one or two points from the original data set and the absolute minimum in the spectra can jump ten or even more days. The general depression around these local minima in the spectra is nevertheless quite stable and this is why we looked for more stable dispersion estimates. We thus invoked in our analysis a fourth type of dispersion estimates.

The dispersion estimate $D_{4,1}^2$ includes in the cross sums *all* pairs whose corresponding time points do not differ more than $\delta$ (decorrelation length). Its modified version $D_{4,2}^2$ additionally weights pairs linearly. The overall number of pairs included in the cross sums is now significantly larger and consequently the corresponding dispersion estimates are statistically more stable. However, the additional inclusion of pairs with longer distances between time points introduces a certain bias into the estimates. We call it *curvature bias* because for linearly increasing or decreasing trends (or trend fragments) this bias is zero.

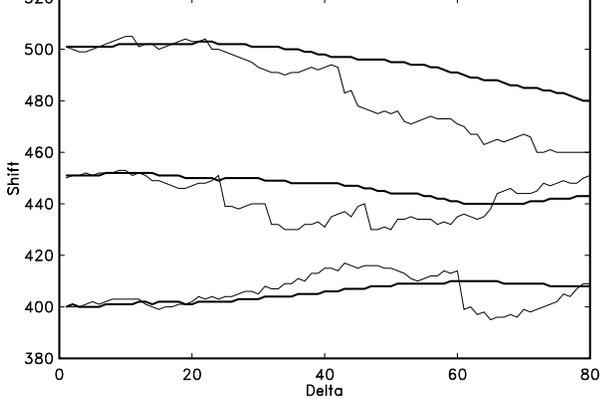
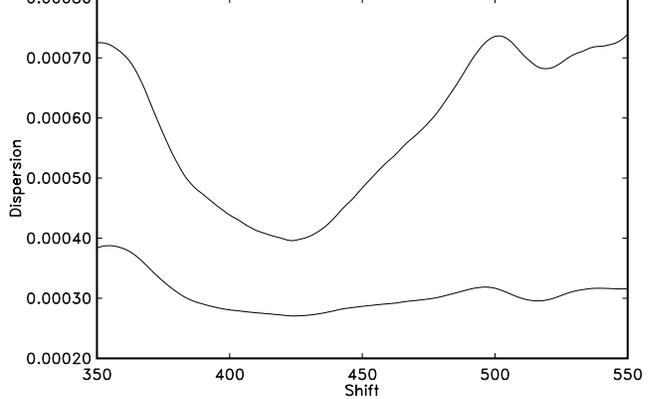

**Fig. 5.** Estimated shifts for three artificial data sets. Thin lines: $D^2_{4,1}$, thick lines: $D^2_{4,2}$

**Fig. 6.** Spectra $D^2_{4,2}$ ($\delta = 20$) for the ST94 data set. Lower curve: all pairs included, upper curve: only A,B pairs included

To investigate the dependence of the estimates $D^2_{4,k}$ on the value of the decorrelation length $\delta$ we generated artificial data sets with known time shifts for the $B$ curve. For this purpose we combined curve $A$ (unshifted) and curve $B$ (shifted by $\tau$), then smoothed the combined curve with an adaptive median filter with length 7, and finally put the smoothed $A$ and $B$ values back into their original places in the data set. In this way the $A$ and $B$ curves can be considered as two replicas of the original source light curve and all the sampling properties and also much of the variability structure of the original data is mirrored in them. In Fig. 5 the results for three different shifts ($\tau = 400, 450, 500$ days) are shown. It can be seen that the first weighting scheme (11) results in much more biased estimates than the second scheme (12). Consequently we preferred to use the second scheme. To balance statistical stability and curvature bias we choose $\delta = 20$, the highest value for which the maximum bias in all of our trial computations did not exceed 2 days. In the following discussion every shift value must therefore be considered with an inherent $\pm 2$ error which originates from the dispersion estimate $D^2_{4,2}$ itself.

In Fig. 6 the resulting dispersion spectra $D^2_{4,2}$, $\delta = 20$, for the ST94 data set are shown. We see that the spectra are now indeed much smoother. The absolute minimum for the spectrum with all pairs included in the cross sums is 424 days and it is 423 days for the spectrum with only $A,B$ pairs included. The bootstrap estimate for the error is $\pm 4$ days. If we take into account the possible bias due to the dispersion estimation scheme (described above) we can formulate our main result:

**The most probable time delay between the two light curves of the double quasar QSO 0957+561 is 423$\pm$6 days**.

It is important to note that the sampling properties of the ST94 data set are good enough to obtain similar shifts for both variants of the statistic $D^2_{4,2}$. Consequently,

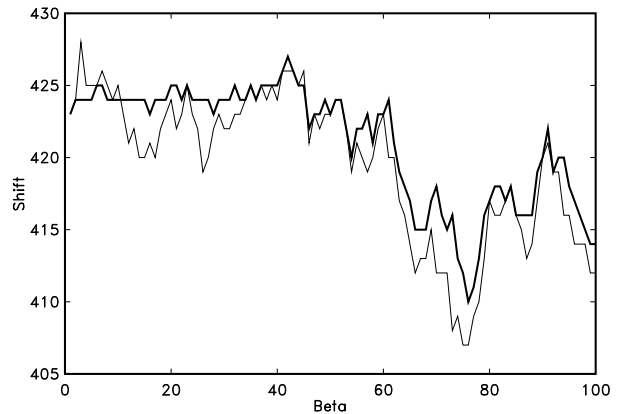

**Fig. 7.** Variability of shift estimates due to changes in the value of $\beta$. Artificial data with known shift 423 days: thick line; original ST94 data set: thin line

it can be safely predicted that similar results will now be also obtained by using the optimal interpolation method of Press et al. (1992a), which is quite sensitive to the amount of regularly spaced gaps in data.

We get additional support for the value 423 days from calculations with the third weighting scheme ($D^2_{4,3}$). In Fig. 7 the dependence of the shift values with minimum dispersion is plotted against the value of the parameter $\beta$. One of the curves is computed for artificial data with a known shift of 423 days (thick line) and one for the original ST94 data set (thin line). As cut-off limit for pair inclusion (to speed up computations) we used a value of $3\beta$.

There is a characteristic "plateau" for values $\beta = 1 \cdots 60$ days in the curve for the artificial data set. If we assume that a significant curvature bias starts for $\beta$ values higher than 60 days we can use dispersion estimates for smaller $\beta$ values as statistical sample with an inner variability just due to the sampling and observational errors. It was extremely satisfying to see that the mean value, the median value and the most frequent value in this sample were all 423 days! However, as can be well seen from Fig. 7, the shift estimation error for particular values of $\beta$ can be as large as 5 days. The bootstrap error for the third weighting scheme is ±4 days. The mean additional error due to the curvature bias is less than 2 days.

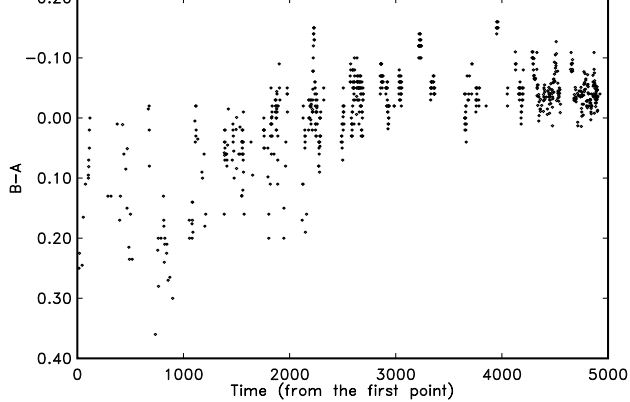

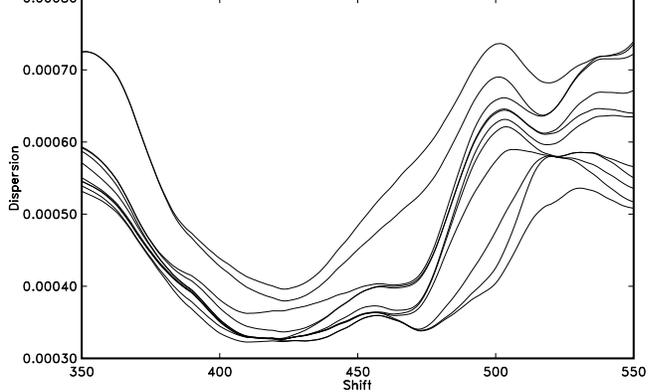

**Fig. 8.** Trend in $B - A$

**Fig. 9.** Spectra $D^2_{4,2}$ for models with polynomial degrees 0 to 10, the level of the dispersion decreases with increasing degrees

### 3.4. Hypothetical microlensing

To check the overall consistency of our solution we proceeded with a rather simple test. For every $B$ value (shifted by $\tau = 423$ days) we looked for the nearest value in the $A$ curve and plotted $B - A$ against time.

In Fig. 8 it can be seen that for the second part of the observational sequence our solution works well, whereas for the first part the situation is quite unclear. First, the dispersion of the observations is significantly higher than for the second part and second there is a clearly visible trend in the differences.

To take into account this trend in the differences between the $A$ and $B$ curves (for the best shift estimate) we introduced a more complicated form for the amplification ratio than $l(t) = l_0$. We introduced simple polynomials with degrees 0 to 10 as models for the *hypothetical microlensing*. The resulting spectra of $D^2_{4,2}$ are shown in Fig.9. The minima and corresponding dispersions for different degrees of polynomials are given in Table 1. In the last column of the table we give the part

$$\nu = 1 - \frac{D^2_{4,2,A,B} - D^2_{4,2,All}}{D^2_{Total}} \qquad (14)$$

of the total dispersion which is explained by the particular model. The total variability of the both light curves (estimated by taking into account weights given by observers) is $D^2_{tot} = 0.00632$.

**Table 1.** Minima for the different degrees of correcting polynomial. Note that several minima are listed for degrees 9 and 10. In the last column the part $\nu$ of the total dispersion which is explained by the particular model is given.

| | A,B | | All | | |
|---|---|---|---|---|---|
| Degree | Shift | Dispersion | Shift | Dispersion | $\nu$ |
| 0 | 423 | 0.000396 | 424 | 0.000271 | 0.980 |
| 1 | 424 | 0.000380 | 424 | 0.000265 | 0.982 |
| 2 | 410 | 0.000362 | 409 | 0.000260 | 0.984 |
| 3 | 423 | 0.000337 | 424 | 0.000252 | 0.987 |
| 4 | 422 | 0.000328 | 422 | 0.000249 | 0.988 |
| 5 | 422 | 0.000328 | 423 | 0.000248 | 0.988 |
| 6 | 422 | 0.000327 | 423 | 0.000248 | 0.988 |
| 7 | 422 | 0.000327 | 423 | 0.000248 | 0.988 |
| 8 | 422 | 0.000327 | 423 | 0.000248 | 0.988 |
| 9 | 431 | 0.000324 | 431 | 0.000247 | 0.988 |
| 9 | 424 | 0.000324 | 424 | 0.000248 | 0.988 |
| 10 | 410 | 0.000323 | 409 | 0.000247 | 0.988 |
| 10 | 423 | 0.000323 | 423 | 0.000248 | 0.988 |
| 10 | 431 | 0.000324 | 431 | 0.000247 | 0.988 |

There are a few peculiar results. For degrees 9 and 10 we listed several minima. The minima around 431 and 410 days are only marginally deeper than the minima around 423 days and they are obviously due to overfitting. Note that we saw these fluctuations already in the simplest spectrum $D^2_2$. The degree 2 gives also a minimum at 410 days which is outside of error brackets for our main result.

To seek an explanation for the additional features in the spectra computed with polynomial perturbations, we proceeded with the following experiment. We decomposed both lightcurves into two by median filtering (see Appendix). The *filtered curves* and the *residual curves* are shown in Fig. 10. We were *extremely* conservative in our filtering process in that we used a short adaptive median filter with a length of 5 points and a gap limit of 5 days. Consequently, the median filtering was applied only to continuous fragments of the light curves, all single and sparsely spaced points were left untouched. In Fig. 11 the $D_{4,2}^2$ type spectra with $\delta = 5$ are shown for the smooth and the residual components. The shorter value for $\delta$ was chosen to show the high frequency fluctuations in the spectra. It is quite clear that the feature around 423 days is mainly due to the variability in the smooth component and that the features around 410 and 431 days are just fluctuations seen in the $D_{4,2}^2$ spectrum for the residuals. Therefore we can attribute the exceptional shift values in Table 1 to the interplay of the under- or overfit and some high frequency details in the original data set. This put aside, it can be said that our main result is quite stable against low frequency perturbations modeled by simple polynomials.

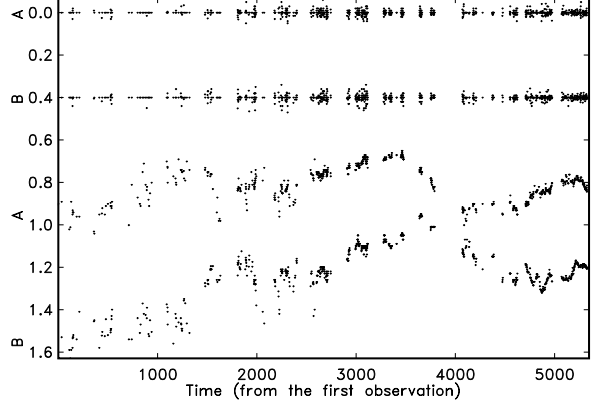

**Fig. 10.** The filtered and residual light curves

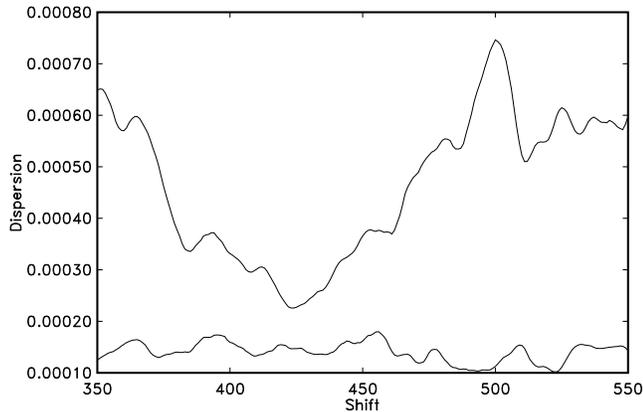

**Fig. 11.** The $D_4^2$ spectra for filtered and residual light curves

was to compute two versions of the $D_{4,2}^2$ statistics – first for all pairs included and the second with only $A,B$ or $B,A$ type pairs included (see Table 1). If our model and our choice of the free parameters in the estimation scheme is correct then the global minima of both dispersion curves must have about the same value. From Fig. 6 we can see that the residual dispersion $D_{4,2}^2$ for the best shift when all pairs were included (0.000271) is quite near to the value of the mean dispersion of the estimated observational errors (0.000251). However, the minimum for the $A,B$ type spectrum (0.000396) is significantly higher. We decreased this value by applying polynomial corrections (up to degree 10) to around 0.000324, but still a significant part of the dispersion remained unexplained. Nonetheless, since the overall variability dispersion for both light curves together is $D^2 = 0.00632$ our models explain $98.0 - 98.8\%$ of the overall variability!

In Paper I we used detrending for **both** light curves to get a self-consistent solution. With the new data the increasing of the correction polynomial degree led to a splitting of the main minimum for degrees higher than 8. Consequently, more complicated models to describe the hypothetical microlensing can be introduced later in our analysis. At this stage it was important to check the general stability of our time delay estimate against different choices of perturbing low frequency components whose possible existence is conjectured theoretically.

The last point can be well illustrated by the simple plot in Fig. 12. We combined the original $A$ curve with the $B$ curve (again shifted by 423 days). We then computed difference values for $A,B$ pairs with distance in time not exceeding 20 days and fitted standard cubic splines with different numbers of knots $(15 \cdots 21)$ into these differences. Depending on the degree of smoothing, different details in the flow of the differences can be detected. The strong depression in the first part of the data span well explains why it was not possible to get the right value for the V89 data *without* previous detrending. Other significant departures from the smooth flow of the differences can be used to explain the residual dispersion. At this stage of the investigation we refrain from more quantitative and precise evaluation of these features.

*3.5. Radio data*

Because the $D_{4,2}^2$ estimator gives more stable spectra we tried it also on the available radio data (Lehár et al. 1992) for 0957+561. As for the optical data, we computed several trial spectra with known shifts to get the optimal value of $\delta$. We found that the value $\delta = 60$ days was the best compromise between resolution and stability. From Fig. 13 it can be seen that the spectra for the new statistic are smoother than that obtained in Paper I, but the instability due to the two (probably outlying) points is still there. It is enough to remove only one or two points

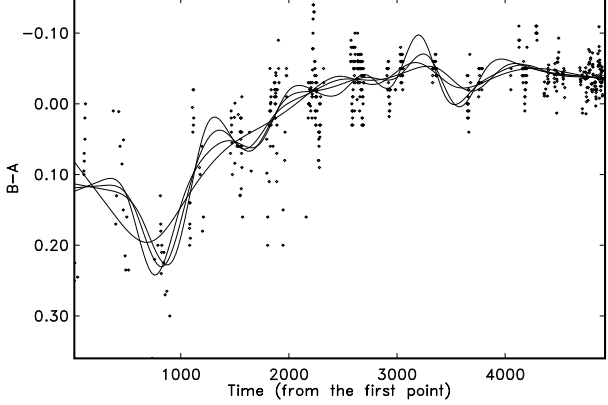

**Fig. 12.** Spline approximations for the differences $B - A$

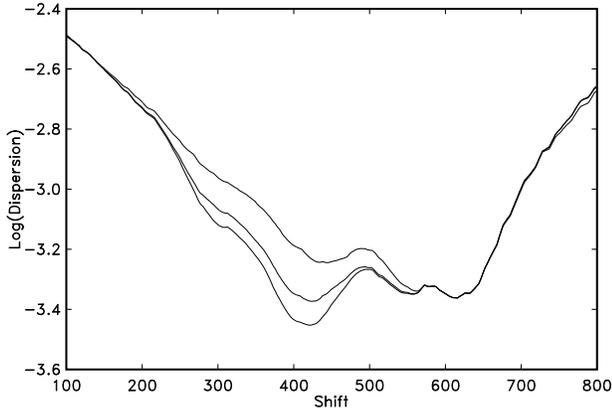

**Fig. 13.** The $D_{4,2}^2$ ($\delta = 60$ days) spectra for the radio data. All data, data with one and two offending points excluded; the removal of each point decreases the dispersion

from the $B$ curve and the minimum shifts significantly. If we look at the spectrum $D_{4,2}^2$ which was computed with the probable outliers removed, we see the local minimum at $421 \pm 25$ days ($D_{4,2}^2 = 0.000353$, error estimate from 1000 bootstrap runs). We used in Paper I the fact that the global minimum for the dispersion can migrate from 533 days to 409 days (after removal of only two points from the $B$ curve) as one of the arguments to stress the compatibility of the radio data with time delays in a wide range and not just around 536 days. We ignored the local minima at 425, 434 and 453 days and did not try to smooth the dispersion curve to get more stable estimates. We see that the smoothed estimator again favors a value around the best estimate found from the optical data. Still, we are not insisting that the radio data give strong support to our best estimate of the delay value from the optical data. There are simply too few points in the published radio data set to compute precise estimates.

the radio data can not be used to compute the exact value for the time delay. However, they are compatible with our result of 423 days.

### 3.6. The composite optical light curve

The obtained value for the time delay and our model for the hypothetical microlensing allows us to build a composite (intrinsic) light curve from both original curves. On Fig. 14 this curve is displayed. To build the composite curve the following parameters were used: time delay 423 days, degree for correction polynomial 7, median filter length 7 days (to obtain a smoother representation), maximum gap length $\delta = 20$ days.

## 4. Discussion

### 4.1. Methodological

Let us first summarize the general methodological principles we tried to follow when analysing the new extensive data set.

- The original data set was treated 'as is', i.e. no censoring, detrending or smoothing was involved.
- We tried to avoid (as far as possible) any interpolations or approximations of the original data.
- We accounted for the significant and fundamental difference between the two kinds of pairs of observations: the ones which contain observations from both light curves and the others where both of the values are from one and the same lightcurve.
- We tried to avoid the introduction of free parameters into our schemes, and if we were forced to do otherwise, we used always several values to see their effect and check the stability of our results against changes of the parameters.
- We avoided any physical interpretation during the data processing stage. This was facilitated by the fact that one of the authors (J.P.) is extremely critical about interpreting the small residual differences as microlensing events.
- Finally we tried to avoid any mathematical models to describe the light curves: no polynomial, Fourier, spline, wavelet, general random process, autoregressive random process or whatever kind of models were involved as models for the original data in different stages of analysis. The only principle we used was the simple recognition that both curves are *statistically continuous*, i.e. nearby observations in time must have nearby values in brightness.

According to these general principles the results we obtained can be ranked. We start from the facts which are methodologically "cleaner" and end with statements which are more open to further discussion and observational follow up.

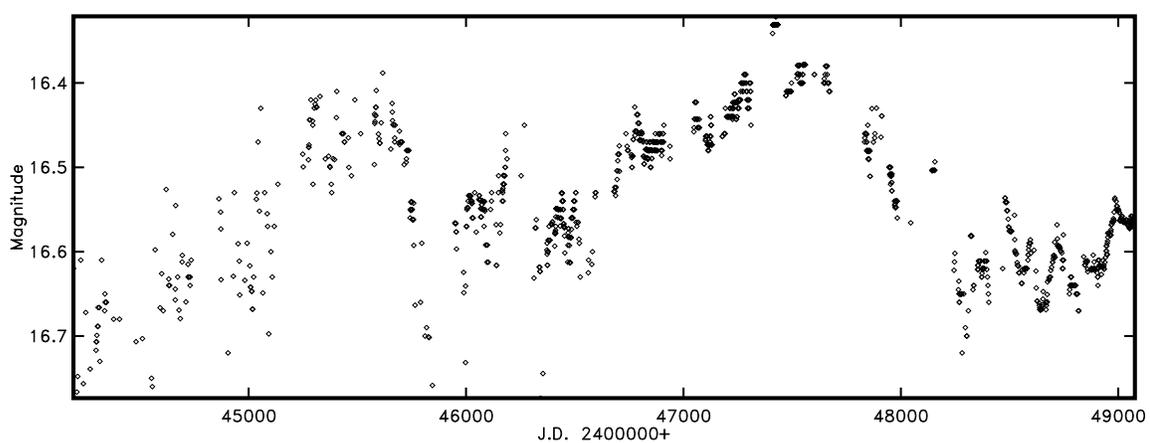

**Fig. 14.** The composite optical light curve of QSO 0957+561

- **The depression in $D^2(\tau)$ around $\tau = 405 \ldots 435$ days is the global minimum.** In all of our calculations with the ST94 data set we were not able to see anything so prominent as the general depression around $405 \cdots 435$ days. There were a lot of other minima (especially in the spectra with short decorrelation length), but they were always less deep and more fluctuation-like than the main minimum. The feature around 536 days which was advocated by Press et al. 1992a is still there but much less apparent.
- **The best stable estimate for the time delay is $423\pm 6$ days.** For $\tau = 423$ days we are able to explain at least 98 percent of the general variability of the two curves. We used different degrees $(0 \cdots 10)$ for the perturbing polynomial $l(t)$, analysed different random subsets of the data set (not all computations are explicitly described in this paper), used different median filter lengths for generating bootstrap samples etc. but were still left with the value 423 days. Even when we ignored the statistical weights (observational errors) given by observers and used unitary weights for all data we got the same results.
- **There is a significant discrepancy between the $A$ and $B$ light curves not fully described by the simple lensing model.** This fact is well demonstrated in Fig.12. We modified our dispersion estimates to take into account smooth (with respect to time) differences. Even after that, some small unexplained variations remained.
- **The previous delay estimates – 409 and 415 days – can be considered as different manifestations of the feature at 423 days.** In Paper I we obtained two estimates for the most probable time delay: $415 \pm 32$ for the optical data and $409 \pm 23$ days for the radio data. Similar results were obtained by other researchers (Vanderriest et al. 1989, Schild et al. 1990, 1993). Our new estimate is well inside the error brackets of the previous estimates, but nevertheless we would like to comment on this specifically. It was interesting to obtain the value around 409 days (see Fig. 11) in our computations with less smoothing ($\delta = 5$ days) or with a high degree correction polynomial (see Table 1). It shows that the previous estimates are just small fluctuations in the bottom of the general depression around the delay of 423 days. These fluctuations are due to the interplay of the small 'bumps' in the $B$ lightcurve and sampling irregularities. Depending on the method used (its degree of smoothing, detrending etc.) these small features show themselves often as local (or sometimes even global) minima in the dispersion spectra.

*4.2. Physical: Microlensing*

The slow trend and the "bump" in the difference $B - A$ (cf. Fig. 12) is most likely due to microlensing by stars in the lensing galaxy. Since, however, the dispersion in the light curves and consequently also in $B - A$ also shows a strong trend, thereby demonstrating the progress in observational technique and experience, we can not completely rule out that the trend in $B - A$ is also a (learning curve like) artefact of the decreasing observational errors.

Taken at face value the variability in the difference $B - A$ is in agreement with numerical simulations concerning its time scale and its brightness change (see, e.g., Kayser (1992), Wambsganss (1993), and references therein). The "bump" can be interpreted as an high amplification event in image $A$ or as a deamplification event due to overfocussing in image $B$ (Kayser et al. 1986).

While one major uncertainty in the determination of the Hubble parameter $H_0$ from the 0957+561 light curves was up to now the controversy on the exact value of the time delay, with our result the remaining uncertainty is due to the mass model of the lensing galaxy and cluster. The mass distribution for this particular object is rather complicated since the main lensing galaxy (at $z = 0.36$) is situated close to the centre of a rich galaxy cluster. Bernstein, Tyson & Kochanek (1993, hereafter BTK) reported the detection of some arcs $20''$ away from the lensing galaxy. They concluded that these arcs can only be explained if a group of galaxies at $z = 0.5$ and $1'.5$ away from the lensing galaxy contributes significantly to the light deflection, thereby complicating the lens models even more. However, Dahle, Maddox & Lilje (1994, hereafter DML) showed that the arcs found by BTK are simple chance alignments of separate objects. Using the method of Kaiser & Squires (1993) DML determined the mass distribution and especially the centre of the galaxy cluster from weak lensing of background galaxies around QSO 0957+561 and thereby refined the model of Kochanek (1991). This model basically uses an elliptical pseudo-isothermal cluster and an elliptical isothermal galaxy.

With our new result for the time delay $\tau = 423$ days we obtain, using the Kochanek-DML model,

$$14 \frac{\text{km}}{\text{s Mpc}} < H_0 < 67 \frac{\text{km}}{\text{s Mpc}}.$$

The remaining uncertainty is due to uncertainty in the mass of the lensing galaxy. Despite this the determination of the time delay for QSO 0957+561 is now already constraining $H_0$ to be smaller than 70 km/(s Mpc) with high statistical significance.

## 5. Conclusions

Using the dispersion estimation technique developed in Paper I in its simplest non-parametric form we were able to show on the basis of the presently available photometric data that the time delay between the images of the double quasar 0957+561 is in the vicinity of 420 days rather than 536 days.

Using refinements to our methods in order to minimise noise due to observational errors and sampling problems we then derived the first accurate and statistical reliable value for the time delay, namely 423±6 days.

The remaining discrepancies between the light curves of the images A and B can most naturally be explained by microlensing due to stars in the lensing galaxy. Using simple polynomial fits we were able to explain 98% of the variability of 0957+561 by our model.

We show again, as in Paper I, that the available radio data are compatible with our result obtained from optical data.

eter is already constrained to be smaller than 70 km/(s Mpc). The remaining uncertainty in the determination of $H_0$ is due to the uncertainty in the mass of the lensing galaxy.

*Acknowledgements.* We are especially grateful to Rudy Schild and David Thomson for providing us with their optical data set prior to publication. Further thanks to P. Helbig for reading the manuscript and helpful discussions. Part of this work was supported by the *Deutsche Forschungsgemeinschaft, DFG* under 436 EST and Schr 417.

## A. Appendix. Adaptive median filtering

Let $f(t_i), i = 1, \ldots, N$ be a time series, $M$ the filter length ($M \ll N$) and $\delta$ the maximum distance between the two observations ($\delta \ll t_n - t_1$) to be considered as nearby. Then for every observation $f(t_i)$ its filtered value $\hat{f}(t_i)$ is defined as the *median* value of the longest subsequence $f(t_{i-n}), n = -L/2, \ldots, L/2, L \leq M$ of the original data set which is centered around the point under discussion and which does not contain gaps longer than $\delta$. In this way single observations remain untouched and groups of nearby observations with less than $M$ observations in them will be filtered with shorter filter lengths. This definition allows us to filter the whole sequence in the best achievable way.

## References

Bernstein G.M., Tyson J.A., Kochanek C.S., 1993, AJ 105, 816 (BTK)
Borgeest U., 1986, ApJ 309, 467
Borgeest U., 1984, A&A 141, 318
Dahle H., Maddox S.J., Lilje P., 1994, ApJ 435, L79 (DML)
Falco E.E., 1992, Lect. Not. Phys. 406, 50
Falco E.E., Gorenstein M.V., Shapiro I.I., 1991, ApJ 372, 364
Kayser R., 1990, ApJ 357, 309
Kayser R., 1992, Lect.Not.Phys. 406, 143
Kayser R., Refsdal S., Stabell R., 1986, A&A 166, 36
Kochanek C.S., 1991, ApJ 382, 58
Lehár J. Hewitt J. N., Roberts D. H., Burke B. F., 1992, ApJ 384, 453
Pelt J., Hoff W., Kayser R., Refsdal S., Schramm T., 1994, A&A 256, 775 (Paper I)
Press W.H., Rybicki G.B., Hewitt J.N., 1992a, ApJ 385, 404
Press W.H., Rybicki G.B., Hewitt J.N., 1992b, ApJ 385, 416
Refsdal S., 1964a, MNRAS 128, 295
Refsdal S., 1964b, MNRAS 128, 307
Schild R.E., 1990, Lect. Not. Phys. 360, 102
Schild R.E., Thomson D.J., 1993, in: *Gravitational Lenses in The Universe*, Proceedings of the 31th Liège International Astrophysical Colloquium, Eds.: J.Surdej et al., p. 415
Schild R.E., Thomson D.J., 1994, preprint (ST94), submitted to AJ
Vanderriest C., Schneider J., Herpe G., Chevreton M., Moles M., Wlérick G., 1989, A&A 215, 1 (V89)
Walsh D., Carswell R.F., Weymann R.J, 1979, Nature 279, 381

Proceedings of the 31th Liège International Astrophysical Colloquium, Eds.: J.Surdej et al., p.369

Young P., 1980, ApJ 241, 507